\documentclass[conference]{IEEEtran}
\IEEEoverridecommandlockouts

\usepackage{cite}
\usepackage{amsmath,amssymb,amsfonts}
\usepackage{algorithmic}
\usepackage{graphicx}
\usepackage{textcomp}
\usepackage{xcolor}
\usepackage{glossaries}
\usepackage{subfig}
\def\BibTeX{{\rm B\kern-.05em{\sc i\kern-.025em b}\kern-.08em
    T\kern-.1667em\lower.7ex\hbox{E}\kern-.125emX}}

\newcommand{\Capacity}{C}

\newcommand{\PowerTx}{P_{\text{t}}}

\newcommand{\NumTxAntenna}{N_{\text{t}}}
\newcommand{\NumSubcarrier}{M}
\newcommand{\CSI}{H}
\newcommand{\SCindex}{k}

\newacronym{SNR}{SNR}{signal-to-noise ratio}
\newacronym{AP}{AP}{access point}
\newacronym{MIMO}{MIMO}{multiple-input multiple-output}
\newacronym{WLAN}{WLAN}{wireless local area network}
\newacronym{RSSI}{RSSI}{received signal strength indicator}
\newacronym{CSI}{CSI}{channel state information}
\newacronym{UCB}{UCB}{upper confidence bound}
\newacronym{UAV}{UAV}{unmanned aerial vehicle}

\begin{document}

\title{A Mechanical Wi-Fi Antenna Device for Automatic Orientation Tuning with Bayesian Optimization\\
\thanks{This work was supported by JSPS KAKENHI Grant Number JP23K26109 and the Telecommunications Advancement Foundation.}
}

\author{\IEEEauthorblockN{1\textsuperscript{st} Akihito Taya}
\IEEEauthorblockA{\textit{Institute of Industorial Science,} \\
\textit{The University of Tokyo}\\
Tokyo, JAPAN \\
taya-a@iis.u-tokyo.ac.jp
}
\and
\IEEEauthorblockN{2\textsuperscript{nd} Yuuki Nishiyama}
\IEEEauthorblockA{\textit{Center for Spatial Information Science,} \\
\textit{The University of Tokyo}\\
Chiba, JAPAN \\
nishiyama@csis.u-tokyo.ac.jp
}
\and
\IEEEauthorblockN{3\textsuperscript{rd} Kaoru Sezaki}
\IEEEauthorblockA{\textit{Center for Spatial Information Science,} \\
\textit{The University of Tokyo}\\
Chiba, JAPAN \\
sezaki@iis.u-tokyo.ac.jp}
}

\maketitle

\begin{abstract}
Wi-Fi access points have been widely deployed in homes, offices, and public spaces.
Some APs allow users to adjust the antenna orientation to improve communication performance by optimizing antenna polarization.
However, it is difficult for non-expert users to determine the optimal orientation, and users often leave the antenna orientation in ineffective positions.
To address this issue, we developed a mechanical Wi-Fi antenna device capable of automatically tuning its orientation.
Experimental results show that antenna orientation could cause a throughput variation of approximately 70\,Mbps under line-of-sight conditions.
Furthermore, Bayesian optimization identified better configurations than random search, demonstrating its effectiveness for orientation tuning.
\end{abstract}


\section{Introduction}
\Glspl{WLAN} have been widely deployed in daily living spaces, where many users connect smartphones to \glspl{AP} for Internet access.
While some \glspl{AP} allow users to adjust the mechanical antenna orientation to improve performance, non-expert users often leave the antennas in ineffective positions.
Because antenna orientation affects polarization and multipath propagation, improper alignment can degrade channel gain and spectral efficiency.
Improving antenna orientation can thus enhance wireless resource utilization and reduce power consumption.

Physical reconfiguration has recently been explored in wireless systems.
For example, many studies have optimized the position and orientation of \glspl{UAV} to improve communication performance \cite{wang2019adaptive,grunblatt2020leveraging}.
These studies suggest that mechanical reconfiguration, not only electronic techniques such as beamforming, can play an important role in performance optimization.

However, little attention has been paid to dynamic mechanical optimization in personal \glspl{AP}, such as home routers.
Even small changes in antenna orientation can significantly affect performance due to fading and polarization.
Therefore, we propose a system that automatically optimizes the mechanical orientation of Wi-Fi antennas using Bayesian optimization.
This approach enables non-expert users to improve communication performance without manual adjustment.

\begin{figure}[!t]
\centering
\subfloat[Implemented device]{\includegraphics[width=0.16\textwidth, clip, trim=10 5 10 5]{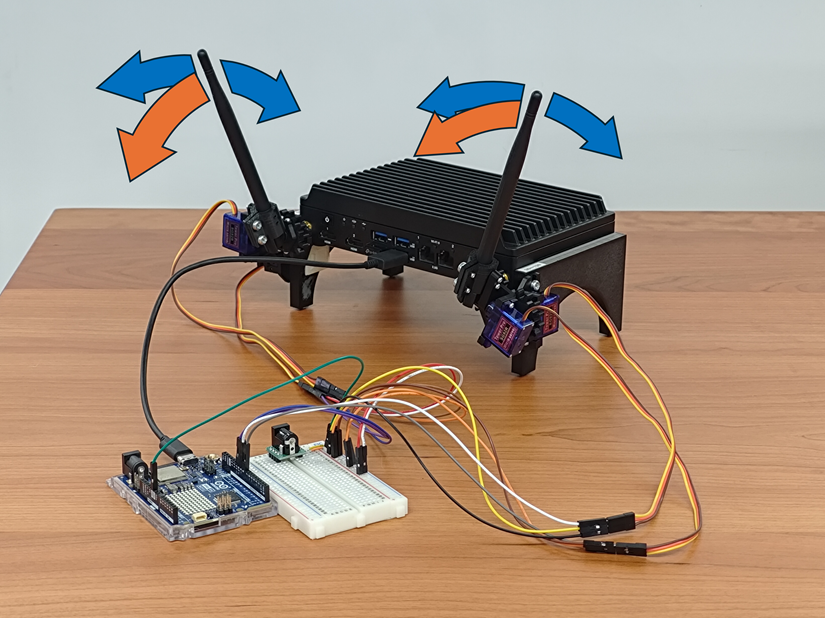}}
\hspace{0.4em}
\subfloat[System Overview]{\includegraphics[width=0.28\textwidth]{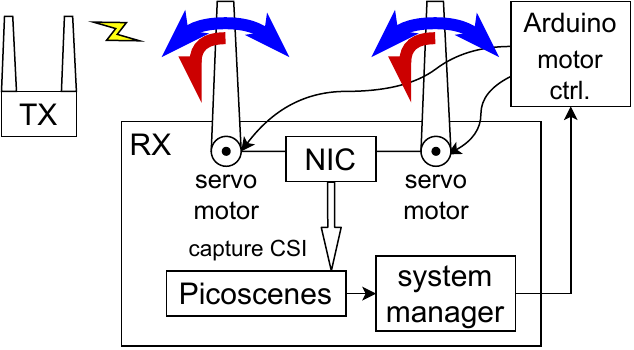}}
\caption{Mechanical Wi-Fi antenna device and its control system. The device can automatically tune the antenna orientation using Bayesian optimization to improve communication performance by calculating channel capacity in real time.}
\label{fig:proposed_device}
\end{figure}

To validate the effectiveness of our concept, we developed a prototype device capable of automatically tuning antenna orientation, as shown in Fig.~\ref{fig:proposed_device}.
The proposed system measures the \gls{RSSI} and \gls{CSI} of the \gls{MIMO} channel and calculates the channel capacity.
Because it is difficult to derive the optimal antenna orientation analytically due to complex radio propagation in indoor environments, we adopt a Bayesian optimization approach that works efficiently with a limited number of measurements.
The system also visualizes the measured \gls{RSSI} and calculated channel capacity in real time, allowing users to intuitively observe how antenna orientation affects communication quality.

Through indoor experiments, we validated that the antenna orientation significantly influences the achievable channel capacity.
Moreover, Bayesian optimization was able to identify better orientations with fewer trials than random search, demonstrating the effectiveness of the proposed system for automatic antenna orientation tuning.

In our demonstration, we will showcase our device that automatically optimizes antenna orientation, along with real-time visualization of \gls{CSI} and channel capacity dynamics.
The automatic tuning process will also be presented to demonstrate the devices' capability to optimize antenna orientation.


\section{System Design}
We designed and implemented a prototype system capable of mechanically controlling the antenna orientation, as illustrated in Fig.~\ref{fig:proposed_device}.
A mini PC was used instead of a commercial \gls{AP} to enable fine-grained \gls{CSI} measurements using Picoscenes~\cite{jiang2021eliminating}.
The prototype system is equipped with two antennas, each of which can be mechanically adjusted in yaw and roll directions.
Two servo motors are attached to each antenna to independently control its yaw and roll angles.
An arduino board is connected to the mini PC via USB, allowing simultaneous motor control and \gls{CSI} collection.

Transmitter sends packets to the receiver periodically by using Picoscenes, and the receiver captures \gls{CSI}.
The system optimizes the antenna orientation to maximize the channel capacity, which reflect the communication quality.
The \gls{MIMO}-OFDM channel capacity $\Capacity$ is calculated as follows \cite{goldsmith2003capacity}:
\begin{align}
    \Capacity = \frac{1}{\NumSubcarrier}\sum_{\SCindex=1}^{\NumSubcarrier} \log_2 \left[ \det \left( I + \frac{\PowerTx}{\sigma^2 \NumTxAntenna} \CSI_\SCindex \CSI_\SCindex^H \right) \right],
\end{align}
where $\NumSubcarrier$, $\NumTxAntenna$, $\PowerTx$, $\sigma^2$, and $\CSI_\SCindex$ denote the number of subcarriers and transmitting antennas, the total transmission power, noise power, and the \gls{CSI} of the $\SCindex$-th subcarrier, respectively.
When the \gls{CSI} is normalized, $\PowerTx/\sigma^2$ corresponds to the average \gls{SNR}, which can be derived from the measured \gls{RSSI}.

Due to the complex characteristics of indoor radio propagation, the optimal antenna orientation cannot be derived analytically.
Therefore, we adopt Bayesian optimization, which efficiently finds near-optimal solutions with a limited number of measurements.
During optimization, the system measures \gls{CSI} over a short periods and uses the averaged channel capacity as the evaluation metric.
The antenna orientation for the next iteration is determined using an \gls{UCB}-based acquisition function.
These steps are repeated until the predefined number of trials is reached.

\section{Evaluations}
We evaluated the proposed system in an indoor environment to validate its effectiveness.
The experiment was designed to confirm that the system can find optimal antenna orientations.
The transmitter (TX) and receiver (RX) were placed facing each other at a distance of 2\,m under line-of-sight (LoS) conditions.
The experiment was conducted in a meeting room containing typical furniture, such as desks and chairs.
Two TX antenna configurations were tested: both antennas were placed in a vertical orientation (V), and both antennas were arranged parallel to each other in a slanted orientation (S), as shown in Figs.~\ref{fig:config_v_tx}~and~\ref{fig:config_s_tx}.

We compared the performance of Bayesian optimization with two baseline search methods: random search and Sobol-sequence-based search.
Fig.~\ref{fig:rslt_convergence} shows the convergence of channel capacity optimization for each method.
The horizontal and vertical axes represent the number of trials and the channel capacity, respectively.
Dots indicate evaluated samples, while solid lines represent the best channel capacity found so far.
As can be seen from the evaluated values, the channel capacity varies by approximately 70\,Mbps depending on the antenna orientation, indicating that mechanical alignment has a substantial impact on communication performance.
The results demonstrate that Bayesian optimization achieved the highest channel capacity in both antenna configurations.

Figs.~\ref{fig:config_v_rx}~and~\ref{fig:config_s_rx} show the antenna orientations identified as optimal by Bayesian optimization.
In both cases, the RX antennas were adjusted to be parallel to the TX antennas, aligning the polarization and improving reception.
This alignment enhances polarization matching, resulting in higher channel gain and throughput.

\def\photowidth{0.11\textwidth}
\begin{figure}[!t]
\centering
\subfloat[TX Orientation (V settings)]{\includegraphics[width=\photowidth]{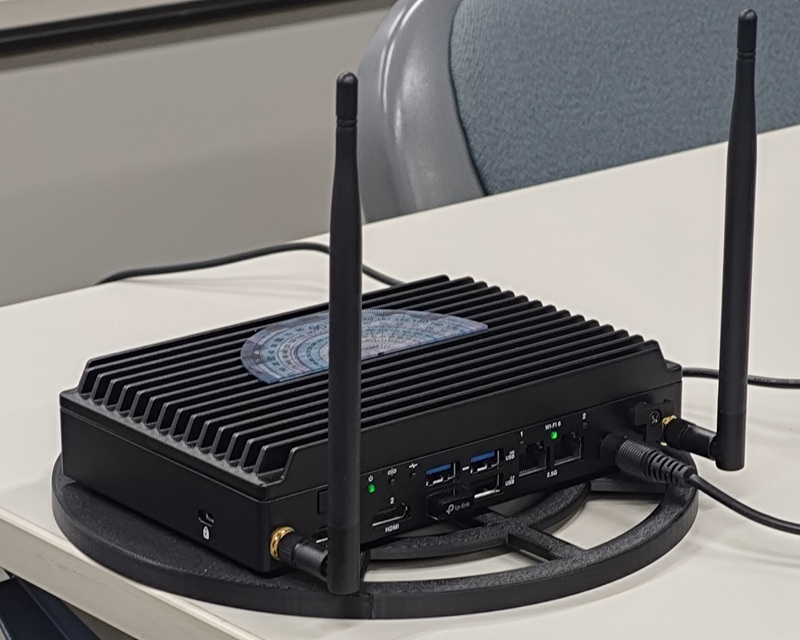}\label{fig:config_v_tx}}
\hspace{0.05em}
\subfloat[RX Results (V settings)]{\includegraphics[width=\photowidth]{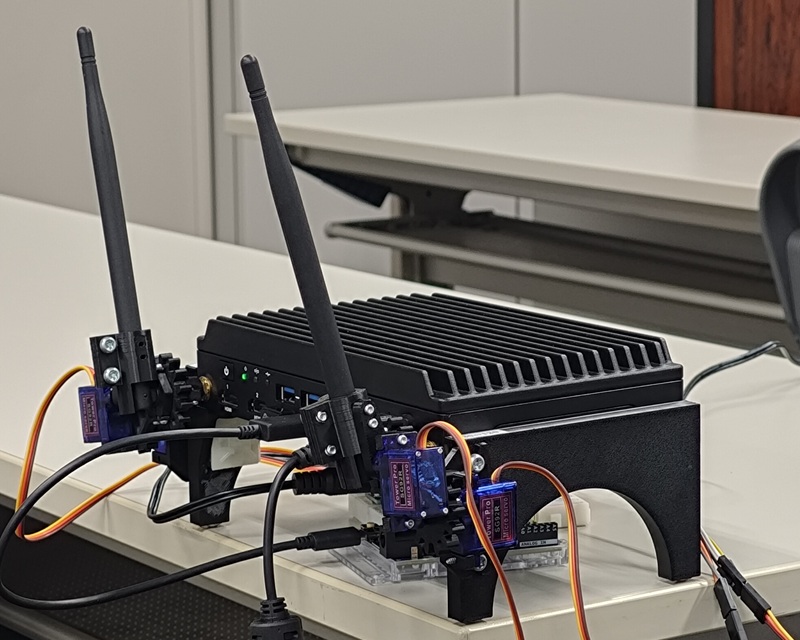}\label{fig:config_v_rx}}
\hspace{0.4em}
\subfloat[TX Orientation (S settings)]{\includegraphics[width=\photowidth]{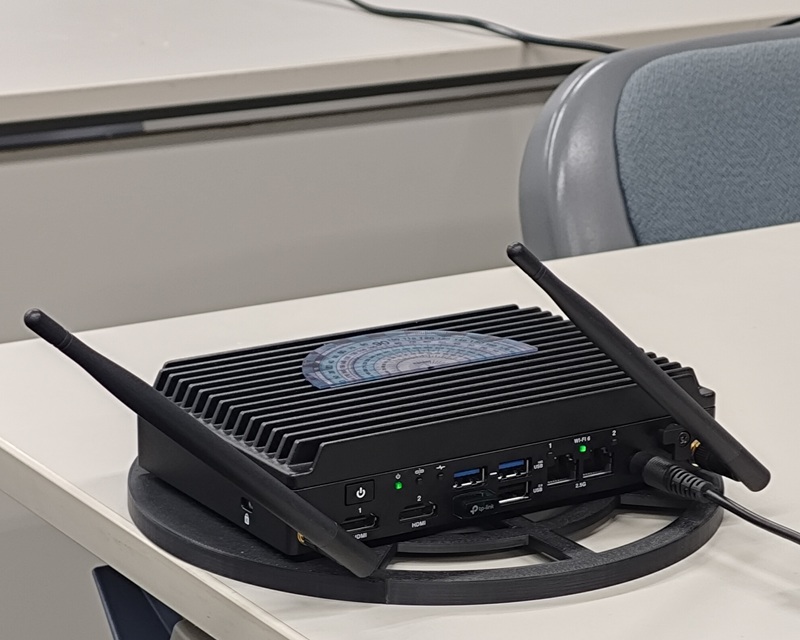}\label{fig:config_s_tx}}
\hspace{0.05em}
\subfloat[RX Results (S settings)]{\includegraphics[width=\photowidth]{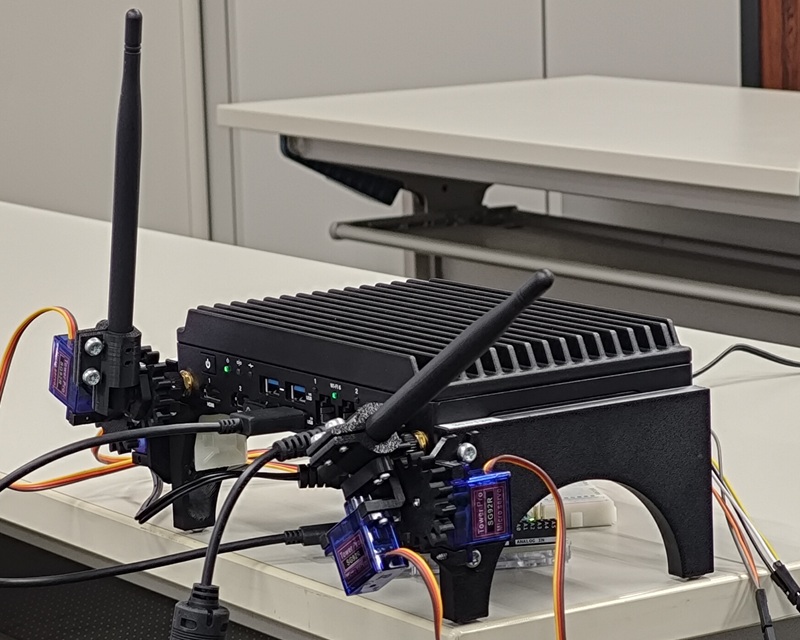}\label{fig:config_s_rx}}
\caption{TX antenna configurations and optimized results of RX antennas. RX antennas were adjusted to be parallel to TX antennas in both configurations.}
\label{fig:antenna_orientations}
\end{figure}

\def\convwidth{0.49\textwidth}
\begin{figure}[!t]
\centering
\includegraphics[width=\convwidth,clip, trim=0 0 0 74]{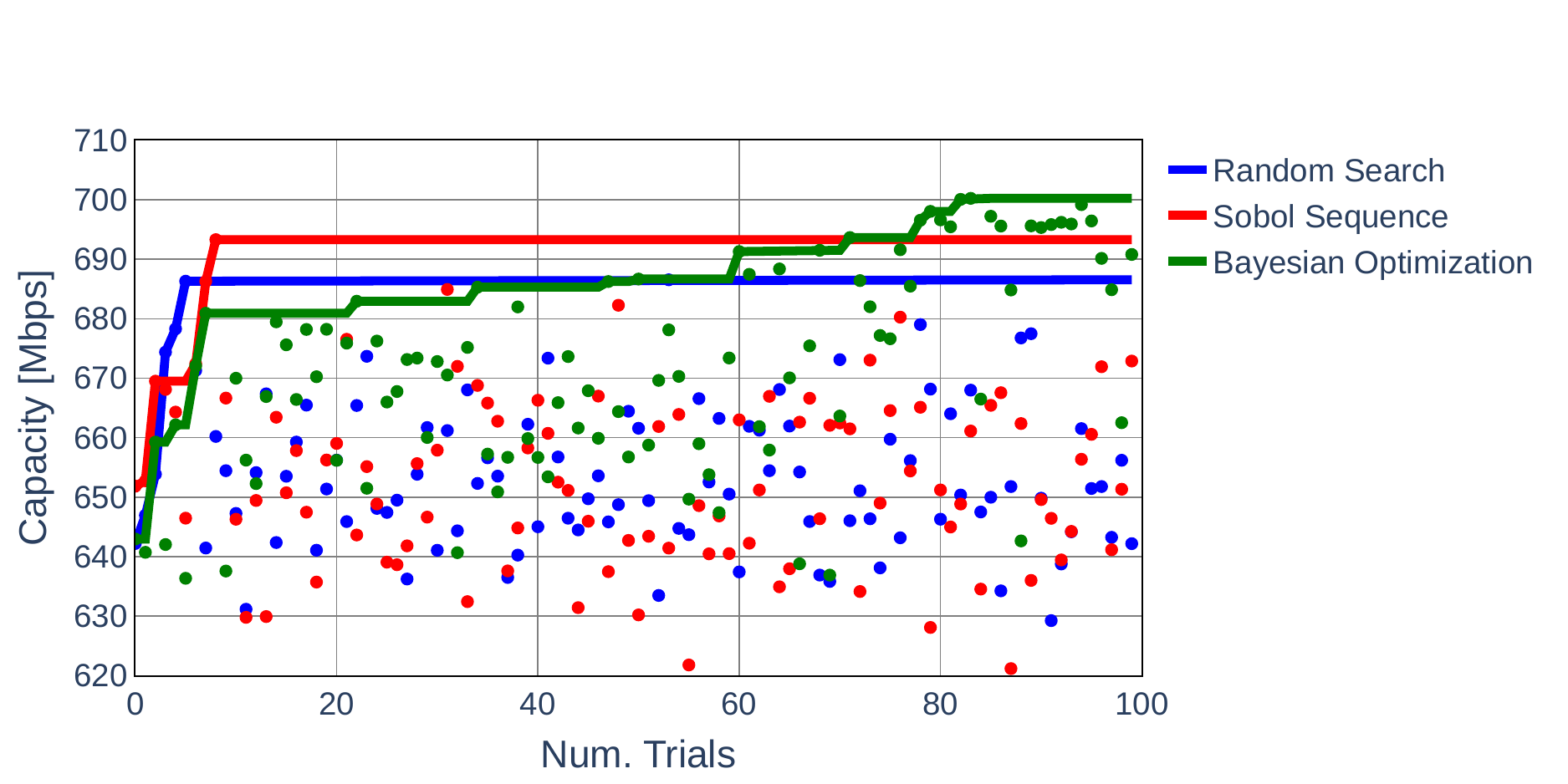}
\caption{Convergence of channel capacity optimization of vertical TX antenna configuration.}
\label{fig:rslt_convergence}
\end{figure}

\section{Conclusion}
This paper presented a mechanical Wi-Fi antenna device and an automatic orientation optimization system based on Bayesian optimization.
Experimental results confirmed that the antenna orientation significantly affects the channel capacity, and that Bayesian optimization efficiently finds near-optimal orientation with fewer trials than random search.
The proposed system successfully identified antenna orientations that maximized polarization alignment and communication performance, demonstrating its practical feasibility for adaptive Wi-Fi optimization.

\bibliographystyle{IEEEtran}
\bibliography{IEEEabrv,myabrv,main}

@STRING{IEEE_J_JSAC       = "{IEEE} J. Sel. Areas Commun."}

@STRING{IEEE_J_JSAC       = "{IEEE} Journal on Selected Areas in Communications"}

@article{goldsmith2003capacity,
  title={Capacity limits of {MIMO} channels},
  author={Goldsmith, Andrea and Jafar, Syed Ali and Jindal, Nihar and Vishwanath, Sriram},
  journal=IEEE_J_JSAC,
  volume={21},
  number={5},
  pages={684--702},
  year={2003},
  publisher={IEEE}
}

@inproceedings{grunblatt2020leveraging,
author = {Gr\"{u}nblatt, R\'{e}my and Gu\'{e}rin Lassous, Isabelle and Simonin, Olivier},
title = {Leveraging Antenna Orientation to Optimize Network Performance of Fleets of {UAVs}},
year = {2020},
doi = {10.1145/3416010.3423225},
booktitle = {Proc. of the 23rd International ACM Conference on Modeling, Analysis and Simulation of Wireless and Mobile Systems},
pages = {253--260},
numpages = {8},
location = {Alicante, Spain}
}

@ARTICLE{wang2019adaptive,
  author={Wang, Zhe and Duan, Lingjie and Zhang, Rui},
  journal=IEEE_J_TWC,
  title={Adaptive Deployment for {UAV}-Aided Communication Networks}, 
  year={2019},
  volume={18},
  number={9},
  pages={4531-4543},
  doi={10.1109/TWC.2019.2926279}
  }

@ARTICLE{jiang2021eliminating,
  author={Jiang, Zhiping and Luan, Tom H. and Ren, Xincheng and Lv, Dongtao and Hao, Han and Wang, Jing and Zhao, Kun and Xi, Wei and Xu, Yueshen and Li, Rui},
  journal=IEEE_J_IOT, 
  title={Eliminating the Barriers: Demystifying {Wi-Fi} Baseband Design and Introducing the {PicoScenes} {Wi-Fi} Sensing Platform}, 
  year={2022},
  volume={9},
  number={6},
  pages={4476-4496},
  doi={10.1109/JIOT.2021.3104666}
}

@STRING{IEEE_J_IOT        = "{IEEE} Internet Things J."}

\end{document}